\documentclass[12pt]{article}
\begin{document}
\vspace*{-.6in}
\thispagestyle{empty}
\begin{flushright}
CALT-68-2204\\
hep-th/9812037
\end{flushright}
\baselineskip = 20pt

\vspace{.5in}
{\Large
\begin{center}
Introduction to M Theory and AdS/CFT Duality\footnote{Work
supported in part by the U.S. Dept. of Energy under Grant No.
DE-FG03-92-ER40701.}
\end{center}}

\begin{center}
John H. Schwarz\\
\emph{California Institute of Technology, Pasadena, CA  91125, USA}
\end{center}
\vspace{1in}

\begin{center}
\textbf{Abstract}
\end{center}
\begin{quotation}
\noindent  
An introductory survey of some of the developments that have taken place
in superstring theory in the past few years is presented. The main focus is 
on three particular dualities. The first one is the
appearance of an 11th dimension in the strong coupling limit of the type IIA
theory, which give rise to M theory. The second one is the duality between
the type IIB theory compactified on a circle and M theory on a two-torus.
The final topic is an introduction to the recently proposed duality between
superstring theory or M theory on certain anti de Sitter space backgrounds and 
conformally invariant quantum field theories.
\end{quotation}

\vfil
\centerline{To be published in the Proceedings of}
\centerline{\it Quantum Aspects of Gauge Theories, Supersymmetry, and Unification}
\centerline{Corfu, Greece -- September 1998}

\newpage

\pagenumbering{arabic}

\section{Introduction}

It is a pleasure to speak in such a remarkable venue -- the old fortress of Corfu City.
It is not often that one gets to meet inside a tourist attraction. The last such occasion
for me was at a conference that took place inside the Chateau de Blois.

Many of the talks in this conference will cover recent developments in superstring
theory and M theory -- in particular, the recent AdS/CFT conjecture. Since the
organizers have chosen to place me first in the schedule, and since not everyone here
is an expert in these matters, I have decided (in consultation with the organizers) to
give a rather general introduction to some of these recent developments. Hopefully,
this will help to provide some of the background that is needed for the
more specialized talks that will follow. It should also relieve those speakers of the
need to give an extended review of the basics.

As I'm sure most of you know, the term M theory was introduced by Witten
to describe the 11-dimensional quantum theory whose low energy effective
description is 11-dimensional supergravity. However, the usage of this term
has been extended by many authors (including myself on occasion) to refer
to the underlying theory that reduces to the five different 10-dimensional
superstring theories in various special limits, as well as the flat 11-dimensional
theory in a sixth special limit. This is somewhat confusing. Therefore,
in a recent talk at the Vancouver conference,
Sen proposed that the term M theory should be reserved for the 11-dimensional
quantum theory and that the (largely unknown) underlying fundamental theory,
which is not specific to any particular spacetime dimension, should be
called {\em U Theory}~\cite{sen2}. He suggested that U could stand for
either ``unknown'' or ``unified''.  This nomenclature makes a lot of sense to me,
so I will try to adhere to it. In the text that follows, I will use the term M theory
and not the term U theory. However,  its usage will occur only in the sense
that Witten originally intended, namely an 11-dimensional quantum theory.

In the first half of this talk (in sections 2 -- 5) I will survey some of the basic
facts about type IIA and type IIB  superstrings in 10 dimensions and the
dualities that related them to M theory. Aside from some minor editing,
these sections are copied from a review that I wrote earlier this year \cite{slansky}.
Then, in section 6 I will give an introduction to the remarkable duality
that has been proposed
between superstring theory or M theory in certain anti de Sitter spacetime
backgrounds and conformally invariant field theories. To be specific, I
will focus on the duality between type IIB superstring theory in an $AdS_5 \times S^5$
background and ${\mathcal N} =4$ supersymmetric gauge theory. As I have 
already indicated, in this talk I will only survey some of the basics, and leave
the discussion of more advanced aspects of this subject to other speakers.
For a more detailed survey of superstring theory and M theory I recommend
the review paper written by Sen \cite{sen98}. Polchinski's new textbook is
also recommended~\cite{polchinski98}. 

\section{Perturbative Superstring Theory}

Superstring theory first achieved widespread acceptance during the {\em first
superstring revolution} in 1984-85.  There were three main developments at
this time.  The first was the discovery of an anomaly cancellation mechanism \cite{green84},
which showed that supersymmetric gauge theories can be consistent in ten dimensions
provided they are coupled to supergravity (as in type I superstring theory)
and the gauge group is either SO(32) or $E_8 \times E_8$. Any other group
necessarily would give uncanceled gauge anomalies and hence inconsistency at
the quantum level.  The second development was the discovery of two new
superstring theories---called {\em heterotic} string theories---with precisely these
gauge groups \cite{gross84}.  The third development was the realization that the $E_8 \times
E_8$ heterotic string theory admits solutions in which six of the space dimensions
form a Calabi--Yau space, and that this results in a 4d effective theory at low
energies with many qualitatively realistic features \cite{candelas85}.  
Unfortunately, there are
very many Calabi--Yau spaces and a whole range of additional choices
that can be made (orbifolds,
Wilson loops, etc.).  Thus there is an enormous variety of possibilities, none of
which stands out as particularly special.

In any case, after the first superstring revolution subsided, we had five
distinct superstring theories with consistent weak coupling perturbation
expansions, each in ten dimensions.  Three of them, the {\em type I} theory and
the two heterotic theories, have ${\mathcal N} = 1$ supersymmetry in the ten-dimensional
sense.  Since the minimal 10d spinor is simultaneously Majorana and Weyl, this
corresponds to 16 conserved supercharges.  The other two theories, called {\em type
IIA} and {\em type IIB}, have ${\mathcal N} = 2$ supersymmetry (32 supercharges)~\cite{green82}.  
In the IIA case the two spinors have opposite handedness so that the spectrum is
left-right symmetric (nonchiral).  In the IIB case the two spinors have the
same handedness and the spectrum is chiral.

The understanding of these five superstring theories was developed in the
ensuing years.  In each case it became clear, and was largely proved, that
there are consistent perturbation expansions of on-shell scattering amplitudes.
In four of the five cases (heterotic and type II) the fundamental strings are
oriented and unbreakable.  As a result, these theories have particularly simple
perturbation expansions.  Specifically, there is a unique Feynman diagram at
each order of the loop expansion.  The Feynman diagrams depict string world sheets, and
therefore they are two-dimensional surfaces.  For these four theories the
unique $L$-loop diagram is a closed orientable
genus-$L$ Riemann surface, which can be visualized
as a sphere with $L$ handles.  External (incoming or outgoing) particles are
represented by $N$ points (or ``punctures'') on the Riemann surface.  A
given diagram represents a well-defined integral of dimension $6L + 2N - 6$.  This
integral has no ultraviolet divergences, even though the spectrum contains
states of arbitrarily high spin (including a massless graviton).  From the
viewpoint of point-particle contributions, string and supersymmetry properties
are responsible for incredible cancellations.  Type I superstrings are
unoriented and breakable.  As a result, the perturbation expansion is more
complicated for this theory, and the various world-sheet diagrams at a given order 
(determined by the Euler number) have to be
combined properly to cancel divergences and anomalies ~\cite{green85}.

\subsection{T Duality}

An important discovery that was made between the two superstring revolutions
is called {\em T duality} \cite{giveon94}.  This is a property of string theories that can be
understood within the context of perturbation theory.  (The discoveries
associated with the {\em second superstring revolution} are mostly nonperturbative.)  T
duality shows that spacetime geometry, as probed by strings, has some
surprising properties (sometimes referred to
as {\em quantum geometry}).  The basic idea can be illustrated by the
simplest example in which one spatial dimension forms a
circle (denoted $S^1$).  Then the ten-dimensional geometry is $R^9 \times S^1$.
 T duality identifies this string compactification with one of a second string
theory also on $R^9 \times S^1$.  However, if the radii of the circles in the
two dual descriptions are denoted $R_1$ and $R_2$, then
\begin{equation}
R_1 R_2 = \alpha'. \label{Tdual}
\end{equation}
Here $\alpha' = \ell_s^2$ is the universal Regge slope parameter, and $\ell_s$
is the fundamental string length scale (for both string theories).  The tension
of a fundamental string is given by
\[
T = 2\pi m_s^2 = \frac{1}{2\pi\alpha'} \, ,
\]
where we have introduced a fundamental string mass scale 
$m_s = (2\pi\ell_s)^{-1}. $

Note that T duality implies that shrinking the circle to zero in one theory
corresponds to decompactification of the dual theory.  Compactification on a
circle of radius $R$ implies that momenta in that direction are quantized, $p =
n/R$.  (These are called {\em Kaluza--Klein excitations}.)  
These momenta appear as masses for states that are
massless from the higher-dimensional viewpoint. String theories
also have a second class of excitations, called {\em winding modes}.  
Namely, a string wound $m$
times around the circle has energy 
\[
E = 2\pi R \cdot m \cdot T = mR/\alpha' \, .
\]
Equation~(\ref{Tdual}) shows that 
the winding modes and Kaluza--Klein excitations are
interchanged under T duality.

What does T duality imply for our five superstring theories?  The IIA and IIB
theories are T dual \cite{ginsparg87}.  
So compactifying the nonchiral IIA theory on a circle of
radius $R$ and letting $R \rightarrow 0$ gives the chiral IIB theory in ten
dimensions!  This means, in particular, that they should not be regarded as
distinct theories.  The radius $R$ is actually a $vev$ of a scalar field, which arises as
an internal component of the 10d metric tensor.  Thus the type IIA and type IIB
theories in 10d are two limiting points in a continuous moduli space of quantum vacua.
The two heterotic theories are also T
dual, though there are technical details involving Wilson loops, which we will not
explain here.  T duality applied to the type I theory gives a dual description,
which is sometimes called I${}^{\prime}$.  The names IA and IB have also been introduced
by some authors.

For the remainder of this paper, we will restrict attention to theories with
maximal supersymmetry (32 conserved supercharges).  This is sufficient to
describe the basic ideas of M theory.  Of course, it suppresses many
fascinating and important issues and discoveries.  In this way we will keep the
presentation from becoming too long or too technical.  The main focus will be
to ask what happens when we go beyond perturbation theory and allow the
coupling strength to become large in the type II theories.  The answer in the
IIA case, as we will see, is that another spatial dimension appears.

\section{M Theory}

In the 1970s and 1980s various supersymmetry and 
supergravity theories were constructed. (See~\cite{salam}, for example.)
In particular, supersymmetry representation theory showed that ten is the largest
spacetime dimension in which there can be a matter theory (with
spins $\leq 1$) in which supersymmetry is realized linearly.  
A realization of this is 10d super Yang--Mills
theory, which has 16 supercharges \cite{brink77}.  
This is a pretty ({\it i.e.}, very symmetrical) classical field theory, but
at the quantum level it is both nonrenormalizable and 
anomalous for any nonabelian gauge group.  However, as
we indicated earlier, both problems can be overcome for suitable gauge groups
(SO(32) or $E_8 \times E_8$) when the Yang--Mills theory 
is embedded in a type I or heterotic string theory.

The largest possible spacetime dimension for a supergravity theory (with spins $\leq 2$),
on the other hand, is eleven.  Eleven-dimensional supergravity, which has 32 conserved
supercharges, was constructed 20 years ago \cite{cremmer78a}.  
It has three kinds of fields---the
graviton field (with 44 polarizations), the gravitino field (with 128
polarizations), and a three-index 
antisymmetric tensor gauge field $C_{\mu\nu\rho}$ (with 84
polarizations).  These massless particles are referred to
collectively as the {\em supergraviton}. 
11d supergravity is also a pretty classical field theory, which has attracted
a lot of attention over the years.  It is not chiral, and therefore not subject
to anomaly problems.\footnote{Unless the spacetime has boundaries.  The
anomaly associated to a 10d boundary can be canceled by introducing $E_8$
supersymmetric gauge theory on the boundary \cite{horava95}.}  
It is also nonrenormalizable, and thus it cannot be a fundamental theory. 
(Though it is difficult to demonstrate explicitly that it is not
finite as a result of ``miraculous'' cancellations, we now know that this
is not the case.)   However, we now
believe that it is a low-energy effective description of M theory, which is a
well-defined quantum theory~\cite{witten95a}.  
This means, in particular, that higher dimension
terms in the effective action for the supergravity fields have uniquely determined
coefficients within the M theory setting, even though they are formally 
infinite 
(and hence undetermined) within the supergravity context.

\subsection{Relation to Type IIA Superstring Theory}

Intriguing connections between type IIA string theory and 11d supergravity have
been known for a long time.  If one 
carries out {\em dimensional reduction} of 11d supergravity
to 10d, one gets type IIA supergravity~\cite{campbell84}.  
In this case dimensional reduction can be viewed as
a compactification on a circle in which one drops all the Kaluza--Klein
excitations.  It is easy to show that this does not break any of the
supersymmetries.

The field equations of 11d supergravity admit a solution that describes a
supermembrane.  This solution has the property that the energy
density is concentrated on a two-dimensional surface.  A 3d world-volume
description of the dynamics of
this supermembrane, quite analogous to the 2d world volume
actions of superstrings, has been constructed~\cite{bergshoeff87}.  
The authors suggested that a
consistent 11d quantum theory might be defined in terms of this membrane, in analogy
to string theories in ten dimensions.\footnote{It is now clear that this cannot be done 
in any straightforward manner, since there is no weak coupling limit in which
the supermembrane describes all the finite-mass excitations.}  
Another striking result was the discovery of
double dimensional reduction~\cite{duff87}.  
This is a dimensional reduction in which one
compactifies on a circle, wraps one dimension of the membrane around the circle
and drops all Kaluza--Klein excitations for both the spacetime theory and the
world-volume theory.  The remarkable fact is that this gives the (previously
known) type IIA superstring world-volume action~\cite{green84b}.

For many years these facts remained unexplained curiosities until they were
reconsidered by Townsend~\cite{townsend95a} and by Witten~\cite{witten95a}.  
The conclusion is that type IIA
superstring theory really does have a circular 11th dimension in addition to
the previously
known ten spacetime dimensions.  This fact was not recognized earlier
because the appearance of the 11th dimension is a nonperturbative phenomenon,
not visible in perturbation theory.

To explain the relation between M theory and type IIA string theory, a good
approach is to identify the parameters that characterize each of them and to
explain how they are related.  Eleven-dimensional supergravity (and hence M
theory, too) has no dimensionless parameters.  As we have seen, there are no
massless scalar fields, whose vevs could give parameters.  The only
parameter  is the 11d Newton constant, which raised to a suitable power 
($-1/9$), gives the 11d Planck mass $m_p$.  
When M theory is compactified on a
circle (so that the spacetime geometry is $R^{10} \times S^1$) another
parameter is the radius $R$ of the circle.

Now consider the parameters of type IIA superstring theory.  They are the
string mass scale $m_s$, introduced earlier, and the dimensionless string
coupling constant $g_s$.  An important fact about all five superstring theories
is that the coupling constant is not an arbitrary parameter.  Rather, it is a
dynamically determined vev of a scalar field, the {\em dilaton,} which is a
supersymmetry partner of the graviton.  With the usual conventions, one has
$g_s = \langle e^\phi\rangle$.

We can identify compactified M theory with type IIA superstring theory by
making the following correspondences:
\begin{equation}\label{M1}
m_s^2 = 2\pi R m_p^3
\end{equation}
\begin{equation}\label{M2}
g_s = 2\pi Rm_s.
\end{equation}
Using these one can derive other equivalent relations, such as
\[
g_s = (2\pi Rm_p)^{3/2}
\]
\[
m_s = g_s^{1/3} m_p.
\]
The latter implies that the 11d Planck length is shorter than the string length
scale at weak coupling by a factor of $(g_s)^{1/3}$.

Conventional string perturbation theory is an expansion in powers of $g_s$ at
fixed $m_s$.  Equation~(\ref{M2}) shows that this is equivalent to an expansion
about $R=0$.  In particular, the strong coupling limit of type IIA superstring
theory corresponds to decompactification of the eleventh dimension, so in a sense M theory
is type IIA string theory at infinite coupling.\footnote{The $E_8 \times E_8$ heterotic
string theory is also eleven-dimensional at strong coupling \cite{horava95}.}
This explains why the eleventh dimension was not discovered in
studies of string perturbation theory.

These relations encode some interesting facts.  The fact relevant to eq.~(\ref{M1})
concerns the interpretation of the fundamental type IIA string.
Earlier we discussed the old notion of double dimensional reduction, which
allowed one to derive the IIA superstring world-sheet action from the 
11d supermembrane (or M2-brane)
world-volume action.  Now we can make a stronger statement:  The fundamental
IIA string actually {\em is} an M2-brane of M theory with one of its dimensions
wrapped around the circular spatial dimension.   No truncation to zero modes is
required. Denoting the string and membrane tensions (energy
per unit volume) by $T_{F1}$ and $T_{M2}$, one deduces that
\begin{equation}
T_{F1} = 2\pi R \, T_{M2}.
\end{equation}
However, $T_{F1} = 2\pi m_s^2$ and $T_{M2} = 2\pi m_p^3$.  Combining these
relations gives eq.~(\ref{M1}). It should be emphasized that all the formulas in this
section are exact, due to the large amount of unbroken supersymmetry.

\subsection{p-Branes and D-Branes}

Type II superstring theories contain a variety of $p$-brane solutions that preserve
half of the 32 supersymmetries. These are solutions in which
the energy  is concentrated on a $p$-dimensional spatial
hypersurface. (Adding the time dimension, the  world volume 
of a $p$-brane has $p+1$ dimensions.)
The corresponding solutions of  supergravity
theories were constructed by Horowitz and Strominger~\cite{horowitz91}.
A large class of these $p$-brane excitations are called
{\em D-branes} (or D$p$-branes when we want to specify the dimension),
 whose tensions are given by~\cite{polchinski95}
\begin{equation} \label{Dtension}
T_{Dp} = 2\pi {m_s^{p+1}}/{g_s}.
\end{equation}
This dependence on the coupling constant is one of the characteristic features
of a D-brane.  It is to be contrasted with the more familiar 
$g^{-2}$ dependence of soliton masses
(e.g., the 't Hooft--Polyakov monopole).
Another characteristic feature of D-branes
is that they carry a charge that couples to a gauge field
in the Ramond-Ramond (RR) sector of the theory. 
(Such fields can be described as bispinors.)
The particular RR gauge fields that 
occur imply that
even values of $p$ occur in the IIA theory and odd values in the IIB theory.

D-branes have a number of special properties, which make them especially
interesting.   By definition, they are branes on which strings can end---D
stands for {\em Dirichlet} boundary conditions.  The end of a string carries a
charge, and the D-brane world-volume theory contains a $U(1)$ gauge field that carries the
associated flux.  When $n$ D$p$-branes are coincident, or parallel and nearly
coincident, the associated $(p + 1)$-dimensional world-volume theory is a
$U(n)$ gauge theory.  The $n^2$ gauge bosons $A_\mu^{ij}$ and their
supersymmetry partners arise as the ground states of oriented strings running
from the $i$th D$p$-brane to the $j$th D$p$-brane.  The diagonal elements,
belonging to the Cartan subalgebra, are massless.  The field
$A_\mu^{ij}$ with $i \not= j$ has a mass proportional to the separation of the
$i$th and $j$th branes.  This separation is described by the vev of a
corresponding scalar field in the world-volume theory.

In particular, the D2-brane of the type IIA theory
corresponds to our friend the supermembrane of M theory, but now
in a background geometry in which one of the transverse dimensions is a circle.
The tensions check, because (using eqs.~(\ref{M1}), (\ref{M2}), and (\ref{Dtension}))
\[
T_{D2} = 2\pi {m_s^3}/{g_s} = 2\pi m_p^3 = T_{M2}.
\]
The mass of the first Kaluza--Klein excitation 
of the 11d supergraviton is $1/R$.  Using eq.~(\ref{M2}),
we see that this can be identified with the D0-brane.
More identifications of this type arise when we consider the magnetic dual of
the M theory supermembrane.  This turns out to be a five-brane, called the
M5-brane.\footnote{In general, the magnetic dual of a $p$-brane in $d$
dimensions is a $(d - p - 4)$-brane.}  Its tension is $T_{M5} = 2\pi m_p^6$.
Wrapping one of its dimensions around the circle gives the D4-brane, with
tension
\[
T_{D4} = 2\pi R \,T_{M5} = 2\pi m_s^5/g_s.
\]
If, on the other hand, the M5-frame is not wrapped around the circle, one
obtains the so-called NS5-brane of the IIA theory with tension
\[
T_{NS5} = T_{M5} = 2\pi m_s^6/g_s^2.
\]
This 5-brane, which is the magnetic dual of the fundamental IIA string, exhibits
the conventional $g^{-2}$ solitonic dependence.

To summarize, type IIA superstring theory is M theory compactified on a circle
of radius $R=g_s \ell_s$.
 M theory is believed to be a well-defined quantum theory in 11d, which is
approximated at low energy by 11d supergravity.  Its 
supersymmetric excitations (which are the only ones known when there is no
compactification) are the
massless supergraviton, the M2-brane, and the M5-brane.  These account both for
the (perturbative) fundamental string of the IIA theory and for many of its
nonperturbative excitations.  The identities presented here are exact,
because they are protected by supersymmetry.

\section{Type IIB Superstring Theory}

In the previous section we discussed type IIA superstring theory and its
relationship to eleven-dimensional M theory.  In this section we consider type
IIB superstring theory, which is the other maximally supersymmetric
string theory with
32 conserved supercharges.  It is also 10-dimensional, but unlike the IIA
theory its two supercharges have the same handedness.  Since the spectrum
contains massless chiral fields, one should check whether there are anomalies
that break the gauge invariances---general coordinate invariance, local Lorentz
invariance, and local supersymmetry.  In fact, the UV finiteness of the 
string theory Feynman
diagrams (and associated {\em modular invariance}) ensures that all anomalies must
cancel.  This was verified from a field theory viewpoint in ref.~\cite{alvarez83}.

The low-energy effective theory that approximates type
IIB superstring theory is type IIB supergravity~\cite{green82,schwarz83}, 
just as 11d supergravity approximates M theory.  In each case the
supergravity theory is only well-defined as a classical field theory, but still
it can teach us a lot.  For example, it can be used to construct $p$-brane
solutions and compute their tensions.  Even though such solutions themselves are
only approximate, supersymmetry ensures that their tensions,
which are related to the kinds of charges they carry, are exact.

\subsection{SL(2,Z) duality}

Another significant fact about type IIB supergravity is that it possesses a
global $SL(2,R)$ symmetry.  It is instructive to consider the bosonic spectrum
and its $SL(2,R)$ transformation properties.  There are two scalar fields---the
dilaton $\phi$ and an {\em axion} $\chi$, which are conveniently combined in a
complex field
\begin{equation}
\rho = \chi + ie^{-\phi}.
\end{equation}
The $SL(2,R)$ symmetry transforms this field nonlinearly:
\begin{equation}
\rho \rightarrow \frac{a\rho + b}{c\rho + d},
\end{equation}
 where $a,b,c,d$ are real numbers satisfying $ad - bc = 1$.  However, in the
quantum string theory this symmetry is broken to the discrete subgroup
$SL(2,Z)$~\cite{hull94}, which means that $a,b,c,d$ are restricted to be integers.  Defining
the vev of the $\rho$ field to be
\begin{equation}
\langle \rho \rangle = \frac{\theta}{2\pi} + \frac{i}{g_s},
\end{equation}
the $SL(2,Z)$ symmetry transformation $\rho \rightarrow \rho + 1$ implies that $\theta$
is an angular coordinate.  More significantly, in the special case $\theta =
0$, the symmetry transformation $\rho \rightarrow - 1/\rho$ takes $g_s
\rightarrow 1/g_s$.  This symmetry, called {\em S duality}, implies that the theory
with coupling constant $g_s$ is equivalent to coupling constant $1/g_s$, so
that the weak coupling expansion and the strong coupling expansion are
identical!

The bosonic spectrum also contains a pair of two-form potentials
$B_{\mu\nu}^{(1)}$ and $B_{\mu\nu}^{(2)}$,\footnote{These are sometimes
denoted $B_{NS}$ and $B_{RR}$.}
which transform as a doublet under
$SL(2,R)$ or $SL(2,Z)$.  In particular, the S duality transformation $\rho
\rightarrow - 1/\rho$ interchanges them.  The remaining bosonic fields are the
graviton and a four-form potential $C_{\mu\nu\rho\lambda}$, with a self-dual
field strength.  They are invariant under $SL(2,R)$ or $SL(2,Z)$.

\subsection{Duality Between Type IIB Superstring Theory and M Theory}

In the introductory section we indicated that the type IIA and type IIB
superstring theories are T dual, meaning that if they are compactified on
circles of radii $R_A$ and $R_B$ one obtains equivalent theories for the
identification $R_AR_B = \ell_s^2$.  Moreover, in sect. 2 we saw that the
type IIA theory is actually M theory compactified on a circle.  The latter fact
encodes nonperturbative information.  It turns out to be very useful to combine
these two facts and to consider the duality between M theory compactified on a
torus $(R^9 \times T^2)$ and type IIB superstring theory compactified on a
circle $(R^9 \times S^1)$.

Recall that a torus can be described as the complex plane modded out by the
equivalence relations $z \sim z + w_1$ and $z \sim z + w_2$.  Up to conformal
equivalence, the periods can be taken to be $1$ and $\tau$, with Im $\tau >
0$.  However, in this characterization $\tau$ and $\tau' = (a\tau + b)/(c\tau +
d)$, where $a,b,c,d$ are integers satisfying $ad - bc = 1$, describe equivalent
tori.  Thus a torus is characterized by a modular parameter $\tau$ and an
$SL(2,Z)$ modular group.  The natural, and correct, conjecture at this point is
that one should identify the modular parameter $\tau$ of the M theory torus
with the parameter $\rho$ that characterizes the 
type IIB vacuum~\cite{schwarz95a,aspinwall95a}!  Then the duality
gives a geometrical explanation of the nonperturbative S duality symmetry of
the IIB theory:  the transformation $\rho \rightarrow - 1/\rho$, which sends
$g_s \rightarrow 1/g_s$ in the IIB theory, corresponds to interchanging the two
cycles of the torus in the M theory description.  To complete the story, we should
relate the area of the M theory torus $(A_M)$ to the radius of the IIB theory
circle $(R_B)$.  The desired formula is a simple consequence of the ones
given above
\begin{equation}
m_p^3 A_M = (2 \pi R_B)^{-1}.
\end{equation}
Thus the limit $R_B \rightarrow 0$, at fixed $\rho$, corresponds to
decompactification of the M theory torus, while preserving its shape.
Conversely, the limit $A_M \rightarrow 0$ corresponds to decompactification of
the IIB theory circle.

The duality can be explored further by matching the various $p$-branes in 9
dimensions that can be obtained from either the M theory or the IIB theory
viewpoints~\cite{schwarz95b}.  
When this is done, one finds that everything matches nicely and
that one deduces various relations among tensions, such as
\begin{equation}
T_{M5} = \frac{1}{2\pi} (T_{M2})^2.
\end{equation}
This relation was used earlier when we asserted that $T_{M2} = 2\pi m_p^3$ and
$T_{M5} = 2\pi m_p^6$.

Even more interesting is the fact that the IIB theory contains an infinite
family of strings labeled by pairs of relatively prime integers $(p,q)$~\cite{schwarz95a}.
These integers correspond to string charges that are sources of the gauge
fields $B_{\mu\nu}^{(1)}$ and $B_{\mu\nu}^{(2)}$.  The $(1,0)$ string can be
identified as the fundamental IIB string, while the $(0,1)$ string is the
D-string.  From this viewpoint, a $(p,q)$ string can be regarded as a bound
state of $p$ fundamental strings and $q$ D-strings~\cite{witten95b}.  
These strings have a very
simple interpretation in the dual M theory description.  They correspond to an
M2-brane with one of its cycles wrapped around a $(p,q)$ cycle of the torus.
The minimal length of such a cycle is proportional to $|p+q \tau|$, and thus
(using $\tau = \rho$) one finds that the tension of a $(p,q)$ string is given by
\begin{equation}
T_{p,q} = 2\pi|p + q\rho| m_s^2. \label{pqtension}
\end{equation}
The normalization has been chosen to give
$T_{1,0} = 2\pi m_s^2$. Then (for $\theta = 0$) $T_{0,1} = 2\pi
m_s^2/g_s$, as expected. 
Note that decay is kinematically forbidden by charge conservation
when $p$ and $q$ are relatively prime.  When they have a common divisor $n$,
the tension is the same as that of an $n$-string system.  Whether or not there
are threshold bound states is a nontrivial dynamical question, which has
different answers in different settings.  In the present case there are no such bound
states, which is why $p$ and $q$ should be relatively prime.

Imagine that you lived in the 9-dimensional world that is described
equivalently as M theory compactified on a torus or as the type IIB superstring theory
compactified on a circle.  Suppose, moreover, you had very high energy
accelerators with which you were going to determine the ``true'' dimension of
spacetime.  Would you conclude that 10 or 11 is the correct answer?  If either
$A_M$ or $R_B$ was very large in Planck units there would be a natural choice,
of course.  But how could you decide otherwise?  The answer is that either
viewpoint is equally valid.  What determines which choice you make is which of
the massless fields you regard as ``internal'' components of the metric tensor
and which ones you regard as matter fields.  Fields that are metric components in one
description correspond to matter fields in the dual one.

\section{The D3-Brane and ${\mathcal N}=4$ Gauge Theory}

The $U(n)$ gauge theory associated with a stack of $n$ D$p$-branes has maximal
supersymmetry (16 supercharges).  The low-energy effective theory, when the
brane separations are small compared to the string scale, is supersymmetric
Yang--Mills theory.  These theories can be constructed by dimensional reduction
of 10d supersymmetric $U(n)$ gauge theory to $p+1$ dimensions.  In fact,
that is how they originally were constructed~\cite{brink77}.  For $p\leq 3$, the low-energy
effective theory is renormalizable and defines a consistent quantum theory.
For $ p = 4,5$ there is good evidence for the existence of nongravitational
quantum theories that reduce to the gauge theory in the infrared.  For $p\geq
6$, it appears that there is no decoupled nongravitational quantum theory~\cite{sen97}.

A case of particular interest, which we shall now focus on, is $p = 3$.  A stack of
$n$ D3-branes in type IIB superstring theory has a decoupled ${\mathcal N} = 4, $ $d = 4$
$U(n)$ gauge theory associated to it.  This gauge theory has a number of special
features.  For one thing, due to boson--fermion cancellations, there are no
$UV$ divergences at any order of perturbation theory.  The beta function
$\beta(g)$ is identically zero, which implies that the theory is scale
invariant (aside from scales introduced by vevs of the scalar fields).  In
fact, ${\mathcal N}=4, $ $d=4$ gauge theories are conformally invariant.  The conformal
invariance combines with the supersymmetry to give a superconformal symmetry,
which contains 32 fermionic generators.  Half are the ordinary linearly
realized supersymmetrics, and half are nonlinearly realized ones associated to
the conformal symmetry.  The name of the superconformal group in this case is
$SU(4|2,2)$.  Another important property of ${\mathcal N}=4$, $ d=4$ gauge theories is
electric-magnetic duality~\cite{montonen77}.  
This extends to an $SL(2,Z)$ group of dualities.
To understand these it is necessary to include a vacuum angle $\theta_{YM}$ and
define a complex coupling
\begin{equation}
\tau = \frac{\theta_{YM}}{2\pi} + i \frac{4\pi}{g_{YM}^2}.
\end{equation}
Under $SL(2,Z)$ transformations this coupling transforms in the usual nonlinear
fashion $\left(\tau \rightarrow \frac{a\tau+b}{c\tau+d}\right)$ and the
electric and magnetic fields transform as a doublet.  Note that the conformal
invariance ensures that $\tau$ is a meaningful scale-independent constant.

Now consider the ${\mathcal N}=4 $ $U(n)$ gauge theory associated to a stack of $n$
D3-branes in type IIB superstring theory.  
There is an obvious identification, that turns out to be correct.
Namely, the $SL(2,Z)$ duality of the gauge theory is induced from that of the
ambient type IIB superstring theory.  In particular, the $\tau$ parameter of the
gauge theory is the value of the complex scalar field $\rho$ of the string
theory.  This makes sense because $\rho$ is constant in the field configuration
associated to a stack of D3-branes.
The D3-branes themselves are invariant under $SL(2,Z)$ transformations.  Only
the parameter $\tau = \rho$ changes, but it is transformed to an equivalent
value.  All other fields, such as $B_{\mu\nu}^{(i)}$, which are not invariant,
vanish in this case.

As we have said, a fundamental $(1,0)$ string can end on a D3-brane.  But by
applying a suitable $SL(2,Z)$ transformation, this configuration is transformed
to one in which a $(p,q)$ string---with $p$ and $q$ relatively prime---ends on
the D3-brane.  The charge on the end of this string describes a dyon with
electric charge $p$ and magnetic $q$, with respect to the appropriate gauge field.  
More generally, for a stack of $n$
D3-branes, any pair can be connected by a $(p,q)$ string.  The mass is
proportional to the length of the string times its tension, which we saw is
proportional to $|p + q\rho|$.  In this way one sees that the electrically
charged particles, described by fundamental fields, belong to infinite
$SL(2,Z)$ multiplets.  The other states are nonperturbative excitations of the
gauge theory.  The field configurations that describe them
preserve half of the supersymmetry.  As a
result their masses saturate a BPS bound and are given exactly by the
considerations described above.

\subsection{Three String Junctions}

An interesting question, whose answer was unknown until recently, is whether
${\mathcal N}=4 $  gauge theories in four dimensions
also admit nonperturbative excitations that preserve
1/4 of the supersymmetry.  To explain the answer, it is necessary to first make
a digression to consider three-string junctions.

As we have seen, type IIB superstring theory contains an infinite multiplet of
strings labeled by a pair of relatively prime integers $(p,q)$.  Three
strings, with charges $(p_i, q_i), $ $i = 1,2,3,$ can meet at a point
provided that charge is conserved~\cite{aharony96,schwarz96}.  This means that
\begin{equation}\label{x}
\sum p_i = \sum q_i = 0,
\end{equation}
if the three strings are all oriented inwards.  
(This is like momentum conservation in an ordinary Feynman diagram.) Such a
configuration is stable, and preserves 1/4 of the ambient supersymmetry
provided that the tensions balance.  It is easy to see how this can be
achieved.  If one regards the plane of the junction as a complex plane and
orients the direction of a $(p,q)$ string by the phase of $p + q\tau$, then
eqs.~(\ref{pqtension}) and (\ref{x}) ensure a force balance.

The three-string junction has an interesting dual M theory interpretation.  If
one of the directions perpendicular to the plane of the junction is taken to be
a circle, then we have a string junction in nine dimensions.  This must have a
dual interpretation in terms of M theory compactified on a torus.  We have
already seen that a $(p,q)$ string corresponds to an M2-brane with one of its
cycles wrapped on a $(p,q)$ cycle of the torus.  So now we join three such
cylindrical membranes together.  Altogether we have a single smooth M2-brane
forming a $Y$, like a junction of pipes.  
The three arms are wrapped on
$(p_i, q_i)$ cycles of the torus.  This is only possible topologically
when eq.~(\ref{x}) is satisfied.

We can now describe a pretty construction of 1/4 BPS states 
in ${\mathcal N}=4$ gauge theory, due to Bergman~\cite{bergman97}.
Such a state is described by a 3-string junction, with the three prongs
terminating on three different D3-branes.  This is only possible for $n \geq
3$, which is a necessary condition for 1/4 BPS states.  The mass of such a
state is given by summing the lengths of each string segment weighted by its
tension.  This gives a result in agreement with the BPS formula.
Clearly this is just the beginning of a long story, since the simple
picture we have described can be generalized to arbitrarily complicated string
webs.  So long as the web is in a plane, charges are conserved at the
junctions, and all string segments are oriented in the way we have described,
the configuration will be 1/4 BPS.  Remarkably, arbitrarily high spins can
occur.  There are simple rules for determining them~\cite{bergman98}.  
There are also related results in ${\mathcal N}=2$ (Seiberg--Witten)
gauge theory~\cite{bergman98b,gaberdiel}.
When the web is
nonplanar, supersymmetry is completely broken, and reliable mass
calculations become difficult.  However, one should still be able to achieve a
reliable qualitative understanding of such excitations.  In general, there are
regions of moduli space in which such nonsupersymmetric states are stable.

\section{Introductory Remarks on AdS/CFT Duality}

Maldacena conjectured a remarkable duality between superstring theory or M
theory in a suitable anti de Sitter space background and conformally
invariant field theories \cite{maldacena97}.  
(Some relevant prior papers
are listed in ref.~\cite{klebanov97}.)
These are dualities in the usual sense: namely, when
one description is weakly coupled the corresponding dual one is strongly
coupled.  Thus, assuming the conjecture is true, it allows the use of
perturbative methods to learn nontrivial facts about the strongly coupled dual
theory.  This subject has developed with breathtaking speed: Maldacena's paper
appeared in November 1997, yet by the Strings 98 conference seven months 
later,\footnote{All the talks (including audio) are available at http://www.itp.ucsb.edu/~strings98/.}
more than half the invited speakers chose to speak on this subject.  What I
propose to do here is to introduce some of the basic ideas of the subject.  I
will not attempt to be very detailed or precise.  

Maldacena arrived at his conjecture by considering the spacetime geometry in the
vicinity of a large number ($N$) of coincident $p$-branes.  The three basic
examples of AdS/CFT duality with maximal supersymmetry are provided by taking
the $p$-branes to be either M2-branes, D3-branes, or M5-branes.  Then the
corresponding world volume theories (in 3, 4, or 6 dimensions) have
superconformal symmetry.  They are conjectured to be dual to M theory or type
IIB superstring in a spacetime geometry that is $AdS_4 \times S^7, AdS_5 \times
S^5$, or $AdS_7 \times S^4$.  The background also has nonvanishing gauge fields
with $N$ units of flux on the sphere.  All three of these solutions to 11d
supergravity or type IIB supergravity were studied over a decade 
ago~\cite{freund,petervn,kim}, but the
duality conjecture is new.

Since I am not trying to be comprehensive, only the case of coincident D3
branes will be described.  However, in order to explain what is special about
the case $p = 3$, we will begin by considering $N$ coincident D$p$-branes.
This is a type IIA configuration if $p$ is even and a type IIB configuration if
$p$ is odd.  As we have discussed, the $(p+1)$-dimensional world-volume theory
(in the infrared) is a maximally supersymmetric $U(N)$ gauge theory. The
low-energy effective action is given by dimensional reduction of supersymmetric
$U(N)$ gauge theory in 10 dimensions.  The coupling constant $g_{YM}$ of such a
gauge theory has dimensions (length)$^{(p - 3)/2}$.  It is related to the
dimensionless string coupling constant $g_s$ of the ambient 10d theory by
\[
g_{YM}^2 = g_s (\ell_s)^{p-3}.
\]
Of course, the dimensionless effective coupling is scale dependent.  At an
energy scale $E$ it is given by
\[
g_{eff}^2 (E) = g_{YM}^2 NE^{p-3}.
\]
Thus perturbation theory applies in the UV for $p<3$ and in the IR for $p > 3$.
 The special case $p = 3$ corresponds to ${\cal N} = 4$ super Yang--Mills
theory in four dimensions, which is known to be a finite, conformally invariant
field theory.

As solutions of type II supergravity, an extremal system of $N$ coincident
D$p$-branes has a string-frame metric
\begin{equation}
ds^2 = f^{-1/2} (ds^2)_d + f^{1/2} (ds^2)_{10-d},
\end{equation}
dilaton
\begin{equation}
e^{2\phi} = g_s^2 (f)^{\frac{3-p}{2}},
\end{equation}
and RR gauge field
\begin{equation}
C_{01\ldots p} = f^{-1} - 1,
\end{equation}
where $d = p + 1$ and
\[
(ds^2)_d = - dt^2 + dx_1^2 + \ldots + dx_p^2\]
\[
(ds^2)_{10-d} = dx_{p+1}^2 + \ldots + dx_9^2 = dr^2 + r^2 d
\Omega_{8-p}^2 
\]
\[
f = 1 + \frac{g_{YM}^2 N}{\ell_s^4 U^{7-p}}
\]
\[
U = r/\ell_s^2 .
\]
The variable $U \sim Tr$ is essentially the energy of a string stretched
between the D-branes at $r = 0$ and a point a distance $r$ from the D-branes.
The surface $r = 0$ is the horizon of this geometry, which can be regarded as
the location  of the D-branes.

The key step in Maldacena's analysis is to isolate the behavior of the
near-horizon geometry by letting $\ell_s \rightarrow 0$ while holding $U$
fixed.  In this limit one finds that
\begin{equation} \label{nearhor}
ds^2 \rightarrow \ell_s^2 \bigg\{ \frac{1}{\sqrt{\lambda}} U^{\frac{7-p}{2}}
(d x^2)_d + \sqrt{\lambda} U^{\frac{p-7}{2}} dU^2
+ \sqrt{\lambda} U^{\frac{p-3}{2}} d\Omega_{8-p}^2 \bigg\},
\end{equation}
where
\begin{equation}
\lambda = g_{YM}^2 N.
\end{equation}
Also,
\[
e^\phi \rightarrow \frac{1}{N} \left[g_{eff} (U)\right]^{\frac{7-p}{2}}.
\]
{}From these equations we see that the string is weakly coupled for $g_{eff}^2
(U) \ll N^{4/(7-p)}$ and that the curvature is of order $[\ell_s^2 g_{eff}
(U)]^{-1}$.  Thus the supergravity approximation is good ({\it i.e.}, stringy effects
are negligible) for $g_{eff}^2 (U) \gg 1$.  Taking $p<6$ and requiring both of
these inequalities gives the requirement $N \gg 1$.  There is much more that
can be said for each specific value of $p$.  However, we will focus on the special case
$p = 3$ from now on, and refer the reader to ref. \cite{itzhaki98} for a discussion of the
other cases. 

Taking $p = 3$, the near-horizon metric in eq. (\ref{nearhor}) simplifies to
\begin{equation}
ds^2 = \ell_s^2 \left\{\frac{1}{\sqrt{\lambda}} U^2 (dx^2)_4 +
\frac{\sqrt{\lambda}}{U^2} dU^2 + \sqrt{\lambda} d\Omega_5^2\right\},
\end{equation}
and the dilaton is constant:
\[
e^\phi = g_s.
\]
Making the change of variables $z = \sqrt{\lambda}/U$, the metric takes the
form
\begin{equation}
ds^2 = \ell_s^2 \sqrt{\lambda} \left\{\frac{(dx^2)_4 + dz^2}{z^2} +
d\Omega_5^2\right\}.
\end{equation}
This describes the product-space geometry $AdS_5 \times S^5$, where both
factors have radius
\[
R = \lambda^{1/4} \ell_s.
\]
We see that stringy effects are suppressed for $\lambda \gg 1$.  Quantum corrections
are small for $N \gg 1$, since the Planck length is given by $\ell_p =
g_s^{1/4} \ell_s$ and $\lambda = g_s N$.  The value of the RR gauge field
corresponds to the self-dual field strength five-form
\[
F_5 \sim N\big(({\rm vol})_{AdS_{5}} + ({\rm vol})_{S^{5}}\big).
\]
In particular the flux $\int_{S^{5}} F_5 \sim N$.  The $AdS_5 \times S^5$
metric has isometries
\[
SO (4,2) \times SO(6) \approx SU(2,2) \times SU(4).
\]
Including the supersymmetries, the complete isometry supergroup is $SU(2,2|4)$.
This contains 32 supercharges transforming as $(4,4) + (\bar 4, \bar 4)$.

\subsection{The Conjecture}

Maldacena's conjecture is that type IIB superstring theory in the $AdS_5 \times
S^5$ background described above is dual to ${\cal N} = 4$, $d=4$ super
Yang--Mills theory with gauge group $SU(N)$.  This is plausible because this theory is
associated to a system of $N$ coincident D3-branes, as we have explained.
The passage from $U(N)$ to $SU(N)$ is a technical detail that I will not attempt to
explain.\footnote{For a discussion of this point, and how it is reconciled
with the SL(2,Z) duality, see ref.~\cite{wittennew}.} 
The duality incorporates the following correspondences:

\begin{itemize}
\item The (large) integer $N$ gives the rank of the gauge group, which
corresponds to the flux of the five-form RR gauge field.

\item The YM coupling constant $g_{YM}$ is related to the string coupling
constant by $g_{YM}^2 = g_s$.  The fact that $g_{YM}$ does not depend on energy
scale corresponds to the fact the dilaton is constant.  

\item The supergroup SU($2,2|4$) is the isometry group of the superstring
theory background and it is the superconformal symmetry group of the ${\cal N}
= 4$ gauge theory.  In the gauge theory, 
16 of the fermionic generators are linearly realized
supersymmetries and the other 16 generate superconformal transformations.

\item The common radius $R$ of the $AdS_5$ and $S^5$ geometries is
related to the `t Hooft parameter $\lambda = g_{YM}^2 N$ of the gauge
theory by $R/\ell_s = \lambda^{1/4}$.
\end{itemize}

As a side remark, let me point out that the $AdS_5 \times S^5$ metric is
conformally flat, because the two radii are equal.  As a result, in addition to
its isometries it has additional conformal isometries.  From the point of view
of the dual gauge theory these correspond to the 24 10-dimensional Lorentz
transformations that are broken by dimensional reduction from $d = 10$ to $d =
4$.  They have no analogs for the M theory backgrounds $AdS_4 \times S^7$ and
$AdS_7 \times S^4$, because the radii are unequal in these cases.

\subsection{The Structure of Anti de Sitter Space}

In the preceding we have presented $AdS_5$ in Poincar\'e  coordinates.  In these
coordinates, $AdS_{d + 1}$ is given by
\begin{equation}
ds^2 = \frac{1}{z^2} \left((dx^2)_d + dz^2\right), \quad z\geq 0.
\end{equation}
The boundary at spatial infinity ($r \rightarrow \infty$) corresponds to $z =
0$.  The AdS/CFT duality is {\em holographic} in the sense that the physics of
the $(d + 1)$-dimensional {\em bulk} is encoded in the $d$-dimensional {\em boundary}
gauge theory.  But how does this hologram work?  The basic idea is that $x^\mu$
coordinates of a point in the bulk correspond to the $x^\mu$ position in the
field theory, whereas the $z = \sqrt{\lambda}/U$ coordinate corresponds to taking the
field theory to have an energy scale (in the Wilsonian sense) $E = U$.

One fact in support of the identification $E \propto U$ is that the bulk
isometry $(x^\mu, z) \rightarrow (a x^\mu, a z)$ corresponds to the field
theory scale transformation $(x^\mu, E) \rightarrow (a x^\mu, E/a)$.  This
argument does not establish the constant of proportionality, however.  An
argument that achieves this is the identification of D-instantons in the bulk
with YM instantons of the gauge theory.  It turns out that the $z$ coordinate
of the D instanton corresponds to the scale size of the YM instanton.  This
suggests the identification $U = \sqrt{\lambda} E$.  It is a somewhat puzzling
fact that this argument gives the factor of $\sqrt{\lambda}$, whereas the
``stretched string'' picture, discussed earlier, does not~\cite{peet98}.

Poincar\'e coordinates do not give a complete description of the Lorentzian
$AdS_{d+1}$ spacetime.  To understand this, it is useful to consider a
hypersurface in $d + 2$ Euclidean dimensions:
\[
x_1^2 + \ldots + x_d^2 - t_1^2 - t_2^2 = - R^2 = - 1.
\]
In the last step the radius has been set equal to one, for convenience.  Next,
we pass to spherical coordinates for both the $x$'s and the $t$'s:
\[
(x_1, \ldots, x_d) \rightarrow  (r, \Omega_p)
\]
\[
(t_1, t_2) \rightarrow  (\tau, \theta).
\]
In these coordinates the hypersurface is $r^2 - \tau^2 = - 1$, and the metric on
this surface is
\begin{equation}
ds^2 = \sum dx_i^2 - \sum dt_j^2
= \frac{dr^2}{1 + r^2} + r^2 d\Omega_p^2 - (1 + r^2) d\theta^2.
\end{equation}
Note that the time-like coordinate $\theta$ is periodic!  This would imply that
the conjugate energy eigenvalues are quantized.  This is definitely not what
type IIB superstring theory on $AdS_5 \times S^5$ gives, so we must pass to the
covering space CAdS.  Strictly speaking, one should speak of ``CAdS/CFT
duality.''  So we replace $\theta \in S^1$ by $t \in R$.  This gives
a global description of the desired spacetime geometry.  Letting $r = \tan
\rho$, the metric becomes
\begin{equation}\label{analog}
ds^2 = \frac{1}{\cos^2 \rho} (d \rho^2 + \sin^2 \rho d \Omega_p^2 - dt^2).
\end{equation}
This has topology $B_{p+1} \times R$ which can be visualized as a solid
cylinder.  The $R$ factor corresponds to the global time coordinate $t$, and
$B_{p+1}$ is the ball interior to $S^p$.  The boundary of the spacetime (at
$\rho = {\pi}/{2}$) is $S^p \times R$.  Thus the spatial coordinates of the
dual gauge theory should be compactified on a sphere $S^p$.  (The conformal
symmetry also requires adjoining the point at infinity.)  The $S^p \times R$
geometry makes the $SO(p+1) \times SO(2)$ subgroup of the $SO(p+1, 2)$ conformal
group manifest.  Because of the conformal invariance, the radius of the $S^p$
is unphysical --- there is no scale.  One recovers Minkowski spacetime by
letting it become infinite.  The relation between the coordinates introduced
here and the Poincar\'e coordinates given earlier is
\[
(z, y^\mu) = \left((t_1 + x_d)^{-1}, t_2 z, x_i z\right).
\]

For many purposes it is useful to consider the ``Euclideanized'' AdS geometry.
This can be obtained by Wick-rotating the $t_2$ coordinate:
\begin{equation}
x_1^2 + \ldots + x_d^2 - t_1^2 + t_2^2 = - 1.
\end{equation}
The symmetry is now $SO(1, d+1)$.  This manifold should not be confused with
Lorentzian signature de Sitter space, which would have $+1$ on
the right-hand side.

As before, this manifold can be described in Poincar\'e coordinates by
\begin{equation}
ds^2 = \frac{1}{z^2} (dz^2 + (dy^2)_d),
\end{equation}
where now
\[
(dy^2)_d = dy_1^2 + \ldots + dy_d^2.
\]
Unlike the Lorentzian case, these coordinates describe the 
space globally. They give a description that is equivalent
to the one given by the metric
\begin{equation}
ds^2 = dr^2 + \sinh^2 r d\Omega_d^2,
\end{equation}
which is the analog of eq. (\ref{analog}).  Another equivalent metric is
\begin{equation}
ds^2 = \frac{4 \sum_{i = 1}^{d + 1} x_i^2}{(1 - \sum x_i^2)^2},
\end{equation}
where $\sum x_i^2 \leq 1$.  The latter form shows that the topology is that
of a ball $B_{d + 1}$ whose boundary is a sphere $S^d$.  Thus the dual
Euclideanized gauge theory should be compactified on a sphere --- $S^4$ for our
main example.  In this case the SO(5) subgroup of the SO(5,1) conformal group is
manifest. Of course, conformal symmetry allows one to let the radius go to
infinity.  

The AdS/CFT conjecture has been made more precise  by Gubser,
Klebanov, and Polyakov \cite{gubser98} and by Witten \cite{witten98a}.  
They give an explicit prescription for relating
correlation functions of the Euclideanized conformal field theory to the bulk
theory path integral for specified boundary behavior of the bulk fields.  I
will not spell out the prescription carefully here, but simply remark that it
requires a one-to-one correspondence of bulk fields $\phi$ and gauge invariant
boundary operators ${\cal O}$.  Denoting boundary values of $\phi$ by $\phi_0$, one
computes the bulk theory path integral with these boundary values $Z(\phi_0)$.
Then this is identified with the correlation function $\langle \exp
\int_{S^{d}} \phi_0 {\cal O}\rangle_{CFT}$.  (I am ignoring lots of technical
details here.)  The requisite correspondences have been verified for large
classes of examples.

The CAdS/CFT duality for Lorentzian signature entails new issues that have been
considered in ref. \cite{kraus98}.  The boundary value problem in this case no longer has
unique solutions, because one can add normalizable (propagating) modes.  The
conclusion of \cite{kraus98}, if I understand it properly, is that nonnormalizable bulk
modes correspond to backgrounds that couple to gauge invariant local operators
of the boundary gauge theory, as in the Euclidean case.  In addition, the
normalizable modes correspond to localized fluctuations of the gauge theory.
The latter do not appear in the Euclidean case, so this
identification goes beyond those proposed in refs. \cite{gubser98,witten98a}.

\subsection{Finite Temperature}

The passage to finite temperature is evident.  One starts with the
Euclideanized theory described above and takes the time coordinate to have
period $\beta$, the inverse temperature.  Fermi fields are required to be
antiperiodic, as usual, which breaks supersymmetry.
The topology of the boundary CFT in this case is $S^p \times S^1$.  Witten
observes that there are two different choices for
the topology of the bulk that would give this boundary \cite{witten98b}.  
The one suggested by
the zero-temperature analysis is $B_{p + 1} \times S^1$ and an alternative
possibility is $S^p \times B_2$.  The first choice corresponds to $AdS_{d + 1}$
at finite temperature and the second to a (Euclidean) AdS-Schwarzschild black
hole.  By comparing the action for the two, Witten argues that there is a phase
transition for $N \rightarrow \infty$.  The low temperature phase exhibits
confinement and a mass gap, whereas the high temperature phase has
deconfinement.  This is roughly the picture one expects for QCD.

\section{Concluding Comments}

I have touched on some of the highlights in the remarkable 
development of superstring theory that has taken place in the past few years.
It continues to amaze me that the rapid pace of progress is being
maintained over such an extended period. Certainly, the implications
of the AdS/CFT duality are still being digested, so it will continue for a while longer.
In the brief introduction to this topic in the preceding section,
I have not mentioned many of the applications and
generalizations that have already been worked out.  
This work already makes it clear that this duality will
teach us a great deal about strongly coupled gauge theories --- in particular
their large N master fields.  It is less clear, but likely, that it will also
enhance our understanding of nonperturbative string theory.
Even so, I have the feeling that qualitatively new insights are still required
to properly address the issue of the cosmological constant and the stabilization of moduli.

\end{document}